\documentclass[aps,prd,twocolumn,amsmath,amssymb,floatfix, longbibliography]{revtex4-1}
\usepackage[dvipsnames]{xcolor}
\usepackage{graphicx}
\usepackage[colorlinks,citecolor=blue,linkcolor=blue,urlcolor=blue,plainpages=false,pdfpagelabels]{hyperref}
\usepackage{bm}

\newcommand{\beq}{\begin{equation}}
\newcommand{\eeq}{\end{equation}}
\newcommand{\bea}{\begin{eqnarray}}
\newcommand{\eea}{\end{eqnarray}}

\begin{document}
\title{Magnetoelectric effect in superconductors with d-wave magnetization}

\author{Alexander A. Zyuzin}
\affiliation{Department of Applied Physics, Aalto University, P. O. Box 15100, FI-00076 AALTO, Finland}

\begin{abstract}
We report on a study of the interplay between the supercurrent and spin-polarization in a two-dimensional superconducting system in the presence of a d-wave symmetric antiferromagnetic exchange interaction (altermagnetism).  
It is demonstrated that the supercurrent exhibits a transverse contribution in the presence of both constant and momentum-dependent exchange interactions. We also discuss the analog of the Edelstein effect for such material, showing that the induced spin polarization is quadratic in the supercurrent and d-wave symmetric.
\end{abstract}

\maketitle

\section{Introduction}
The magnetoelectrics in non-centrosymmetric superconductors \cite{Edelstein_1} has recently gathered substantial attention particularly in its experimental application in the non-reciprocal superconducting response \cite{Edelstein_2}, as discussed, for example, in a recent review \cite{Review_diode}.
Specifically, the Edelstein magnetoelectric effect is the generation of spin polarization induced by an applied supercurrent, while its inverse scenario is the diode-like effect, i.e. the critical current is different for
two opposite directions, generated in the presence of an external magnetic field. One of the underlying reasons for these phenomena is the violation of spatial inversion symmetry caused by the spin-orbit interaction or the inhomogeneous magnetic exchange field acting on a momentum-dependent spin splitting of the energy bands  \cite{Rashba_Pekar, Ogg_1966, IvchenkoKiselevPTS1992}, all giving rise to the coupling between electron spin polarization and charge current \cite{Dyakonov_book}. 

In this paper, we consider a case of centrosymmetric metals hosting a collinear antiferromagnetic (AFM) order parameter with a d-wave symmetry, \cite{Hayami_2019, Hariki_PRB2019, Zunger_PRB2020, Evarestov_Chem2022}. 
Such AFM order induces a specific d-wave momentum-dependent spin-splitting of the Fermi surface of conducting fermions \cite{Rashba_Pekar, Ogg_1966, IvchenkoKiselevPTS1992}.  
The extended symmetry classification of anisotropic magnetic order has been recently reviewed in \cite{Smejkal_PRX2022_landscape, Smejkal_PRX2022, Jungwirth_PRX2022}. 
The representative materials displaying this feature include, for example, collinear-type AFMs: metallic $\textrm{RuO}_2$, $\textrm{Mn}_5\textrm{Si}_3$,  $\textrm{VNb}_3\textrm{S}_6$, semiconducting $\textrm{MnTe}$, and many more \cite{ Feng_NatElec2022, Bose_NatElec2022, Smejkal_PRX2022_landscape, Smejkal_PRX2022, Jungwirth_PRX2022, Gonzalez_PRL2023}.
Furthermore, strain-stabilized superconductivity has been recently observed in thin films $\textrm{RuO}_2$ with $T_c\approx 1.8K$ depending on the film thickness, \cite{Uchida_PRL2020, Ruf_Nature2021, Occhialini_PRM2022}.
Motivated by the recent experimental progress, the theoretical investigation of the d-wave AFM exchange coupling on superconductivity became an intensive area of research, including the study of Andreev reflection and Josephson current \cite{Linder_PRB2023, Papaj_PRB2023, Linder_PRL2023, Neupert_arxiv2023, Beenakker_PRB2023}, inhomogeneous states in a d-wave superconductor with d-wave AFM \cite{Schaffer_arxiv2023}, or even exotic Majorana modes \cite{Cano_arxiv2022}. For a recent highlight article, see \cite{Mazin_arxiv2022}. 

In this context, the question of superconductivity and magnetoelectrics comes up naturally. 
Clearly, in the centrosymmetric superconductors with d-wave magnetization, in contrast to the Edelstein effect in polar superconductors, the induced spin polarization of carriers is proportional to the even power of supercurrent and exhibits a d-wave symmetry. Application of a constant exchange field leads to a transverse supercurrent response.

\section{Model}
The Bogoliubov - deGennes Hamiltonian, describing clean two-dimensional superconducting material subject to an isotropic magnetic exchange (or the Zeeman effect of a magnetic field) and d-wave AFM exchange interactions, is given by
\begin{subequations}
\begin{align}\label{Hamiltonian11}
&H_{\rm BdG}(\bm{k}) = 
 \left( \begin{matrix}
 H_0(\bm{k}) - \mu &\Delta \\
\Delta^* & -\sigma_y H_0^*(-\bm{k})\sigma_y + \mu
 \end{matrix}\right),
 \\\label{Hamiltonian12}
&H_0(\bm{k}) = \frac{\bm{k}^2}{2m} + \beta k_x k_y \sigma_z + \gamma (k_x^2- k_y^2) \sigma_z+ \bm{h} \cdot \boldsymbol{\sigma},
\end{align}
\end{subequations}
where $\bm{k}=(k_x,k_y)$ and $m$ are the momentum and mass of electrons, $\mu$ is the chemical potential, $\Delta$ is the s-wave superconducting gap, and $\boldsymbol{\sigma} = (\sigma_x,\sigma_y,\sigma_z)$ is a vector of Pauli matrices in spin space. We will use $\hbar=k_{\mathrm{B}}=1$ units henceforth. 
The second and third terms in (\ref{Hamiltonian12}) describe the d-wave exchange interaction (dubbed "altermagnetism" in some literature \footnote{We will refrain from using altermagnetism and adopt conventional notations of collinear antiferromagnets}) characterized by a parameters $\beta$ and $\gamma$, while the forth term describes the Zeeman (or constant exchange) field spin-splitting $\bm{h} = (h_x, h_y, h_z)$, which can have both in-plane and out-of-plane components.  
We assume that $h$ does not affect the AFM order.

Although we consider a specific $d_{xy}-d_{x^2-y^2}$ wave symmetry field for metals with two-dimensional Fermi surface, the results can be readily extended to situations including more general combinations of the $(d, g, i)$-wave symmetry terms \cite{Smejkal_PRX2022} and to higher dimension as well.

Here will consider a regime of weak exchange, $m |\beta|< 1$, $m |\gamma|< 1$ and $|h|<\mu$. The system tuned to a quadratic band-touching point van-Hove singularity $m|\beta| > 1$, $m|\gamma| > 1$ will be studied separately. 
%As the theoretical investigation of superconducting systems with d-wave exchange interaction has been initiated only recently, it is useful to explore some of their basic properties before going into the studies of magnetoelectric effect. 
It is useful to explore the basic properties of superconducting systems with d-wave exchange interaction before going into the studies of magnetoelectric effect.

Initially, one must distinguish between various manifestations of superconducting correlations in the system. 
Namely, one can realize a scenario in which the superconducting gap in the material is induced by the proximity effect from other superconductor. In this case, $\Delta$ is a proximity induced minigap which might be treated as a model parameter depending on the delicate interplay between the material properties and the contact transparency. On the other hand, one can consider a situation in which the superconducting gap is intrinsic. Let us first revisit the former case. 

%%%%%%%%%%%%%%%%%%%
\begin{figure}[t]
\centering
\includegraphics[width=4.16cm]{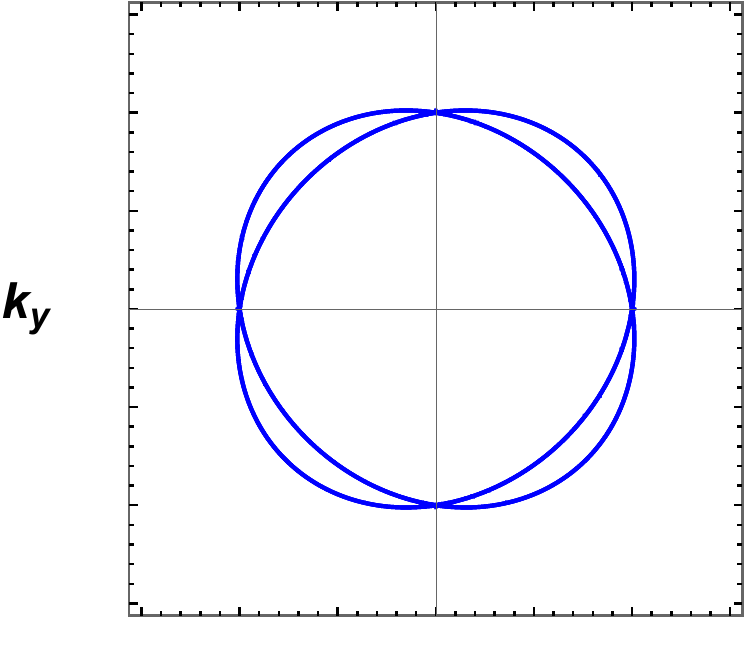} \includegraphics[width=3.74cm]{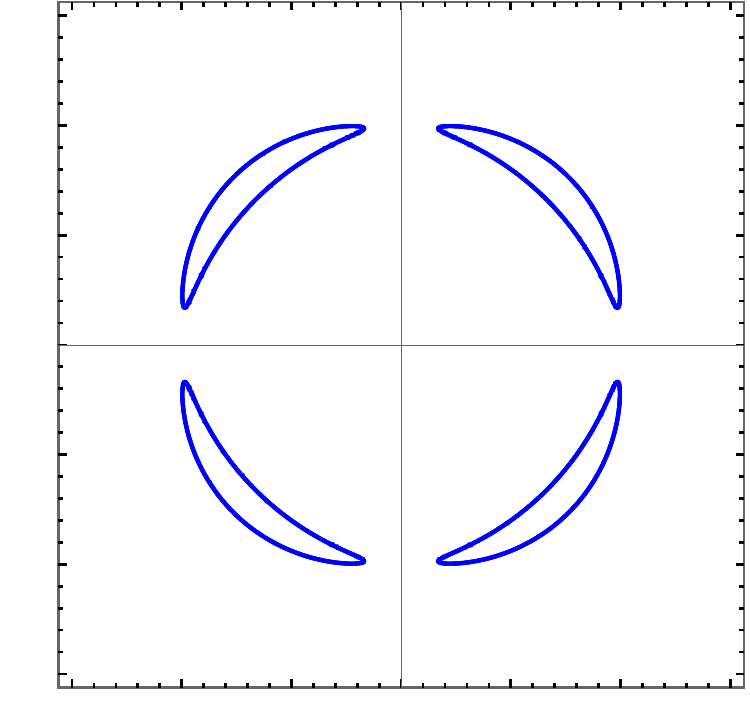}\\
\includegraphics[width=4.16cm]{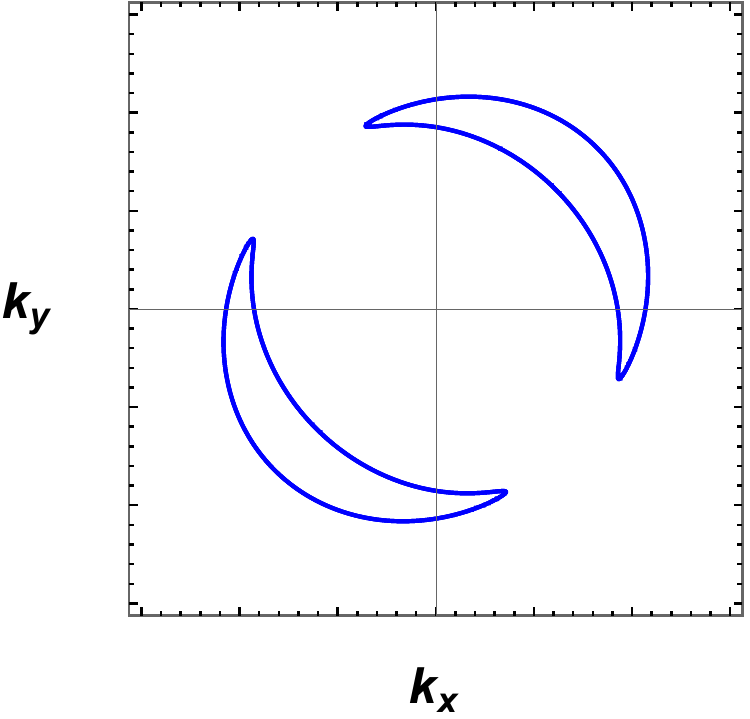} \includegraphics[width=3.74cm]{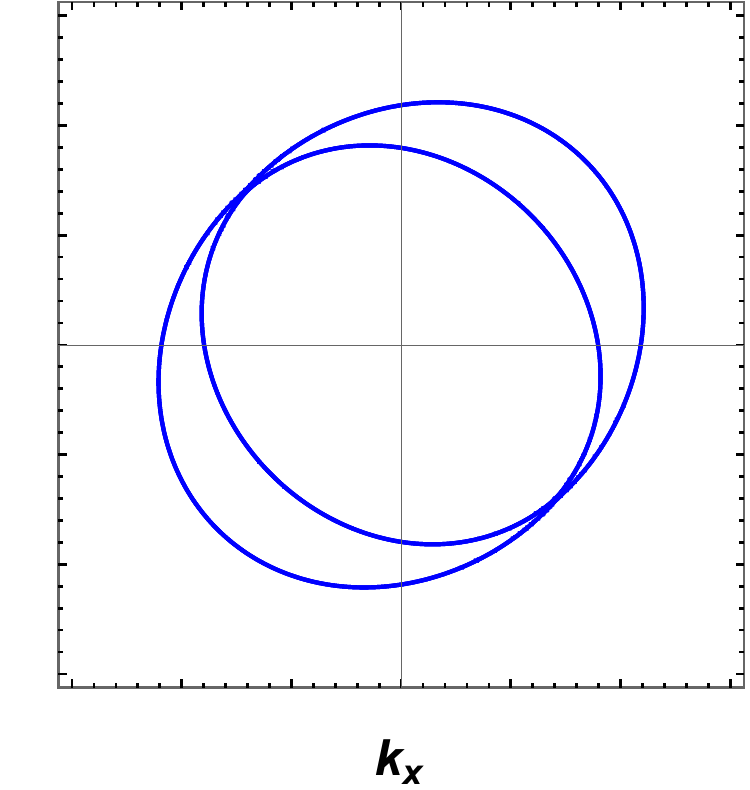}\\
\caption{\label{fig1} Dispersion relation $E_{\pm, s}({\bm k})=0$ for the two-dimensional d-wave exchange field with proximity induced superconducting minigap $\Delta$ at ${\bm h}=(0,0,h_z)$ and $\gamma=0$. (a). At $h_z=0$, $\beta\neq 0$ and $\Delta=0$. (b). At $h_z=0$, $\beta\neq 0$, turning on $\Delta \ll m |\beta| \mu$, gaps out certain regions of the Fermi surface until at $\Delta = 2m |\beta| \mu$ it becomes all gapped. (c) and (d). Now at $\beta\neq 0$ and $\Delta \ll m |\beta| \mu$, the increase of the $h_z$-term closes the gap and tend to spin-split the Fermi surface isotropically.}
\end{figure}
%%%%%%%%%%%%%%%%%

Diagonalizing the Hamiltonian (\ref{Hamiltonian11}), one obtains the following quasiparticle dispersion:
$
E_{\pm, s}({\bm k}) =  \pm \sqrt{ \xi_{\bm k}^2+|\Delta|^2} +s \{h_x^2+h_y^2 + [h_z + \beta k_x k_y+\gamma(k_x^2-k_y^2)]^2\}^{1/2}
$, where $s=\pm 1$ denotes the exchange field induced band splitting and $ \xi_{\bm k} = {\bm k}^2/(2m) -\mu $. Keeping $\Delta$ as a parameter, one observes that the application of the exchange field results in the emergence of nodes within the gap function. Setting $h=0$, $\gamma=0$, one finds the existence of nodes provided the following inequality holds $|\sin(2\phi)| \geqslant \frac{1}{m|\beta|}\frac{|\Delta|}{\sqrt{|\Delta|^2+\mu^2}}$, where $\phi$ is the momentum angular coordinate. With the increase of parameter $|\beta|$, the gap closes at four points on the $k_x-k_y$ plane with coordinates determined by $|\sin(2\phi)| = \frac{1}{m|\beta|}\frac{|\Delta|}{\sqrt{|\Delta|^2+\mu^2}}$ and $k=\{2m\mu(1+|\Delta|^2/\mu^2)\}^{1/2}$. With the further increase of $|\beta|$, each nodal points transform into two Fermi arcs. The application of $h_z$ eliminates one pair of Fermi arcs stretching the other pair. The effect of $h_x$-term is to close the gap isotropically, while the effect of $\gamma$ is to rotate the plot around the centre of origin. The evolution of the gap nodes as a function of the $\beta$ and $h_z$ is illustrated in Fig. (\ref{fig1}). All in all, the $\beta, \gamma$ and $h$ terms break time-reversal symmetry, therefore, contributing to the vanishing of the gap function $\Delta$.

%%%%%%%%%%%%%%%%%%%
\begin{figure}[t]
\centering
\includegraphics[width=3.85cm]{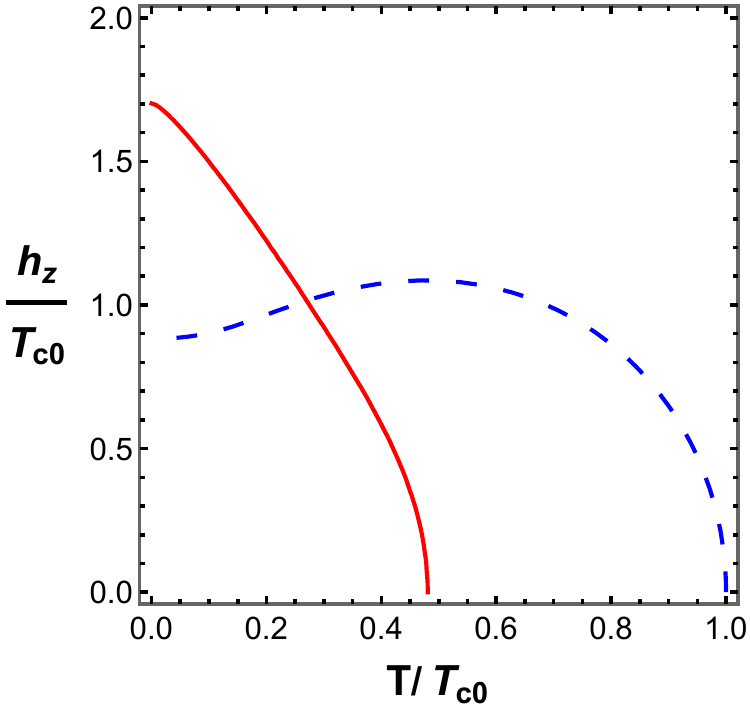} \includegraphics[width=4.05cm]{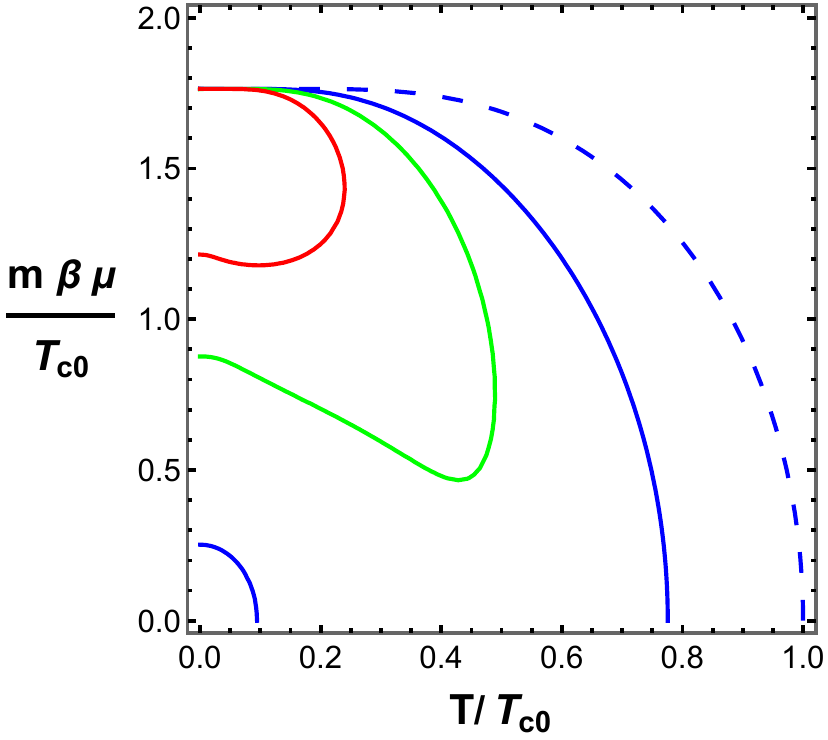}
\caption{\label{fig2} The phase transition curves between the BCS and the normal metal states are shown in two scenarios ($\gamma=0$). (a) Fixed parameter $m\beta\mu/T_{c0} = (0, 0.7)$,  where $T_{c0}$ is the superconducting transition temperature at $\beta=0$ and $h_z=0$. The d-wave term tends to increase the transition temperature at large $h_z$. (b) Fixed parameter $h_z/T_{c0} = (0, 0.9, 1.1, 1.3)$. The increase of the d-wave term gives rise to a superconducting pocket at large $h_z$.}
\end{figure}
%%%%%%%%%%%%%%%%%

To get more insight into superconducting correlations, it is helpful to examine the spatial dependence of the Cooper pair wave function. It can be analyzed by exploring the kernel in particle - particle ladder, $\Pi({\bm r}) =  T\sum_{n}\mathrm{tr}_2  G(\omega_n,{\bm r}) [G(-\omega_n,{\bm r})|_{{\bf h},\beta,\gamma \rightarrow -{\bf h},-\beta,-\gamma}]$, where $\mathrm{tr}_2$ is the trace over the spin Pauli matrices. 
At $\gamma=0$ and $h_x=h_y=0$, without loosing the generality, the electron Green function in spatial coordinate representation, $G(\omega_n, {\bm r}) = \int \frac{d^2k}{(2\pi)^2}[i\omega_n +\mu -H_0(\bm{k}) ]^{-1}e^{-i {\bm k}\cdot{\bm r}}$, is given by
\begin{align}\nonumber
&G(\omega_n, {\bm r}) = - \frac{m}{2\pi}\sum_{s=\pm1} \frac{1+s\sigma_z}{\sqrt{1-\beta^2m^2}} \\
&\times K_0\left(-i \mathrm{sgn}(\omega_n) k_{\mathrm{F}}r f_s(\phi_r) \sqrt{1+\frac{i\omega_n-s h_z}{\mu}} \right),
\end{align}
where $K_0(z)$ is the modified Bessel function, $k_{\mathrm{F}} = \sqrt{2m\mu}$ is the Fermi momentum, $\omega_n = (2n+1)\pi T$ with $n \in \mathbb{Z}$ is the Matsubara frequency at temperature $T$,  and $f_s(\phi_r) = \sqrt{\frac{1- s \beta m \sin(2\phi_r)}{1-\beta^2m^2}}$ is introduced for brevity with $\phi_r$ being the spatial coordinate azimuth angle. Using asymptotic expansion $K_0(z)\approx \sqrt{\frac{\pi}{2z}}e^{-z}$ at $|z|\gg 1$, one obtains
\begin{align}\label{correlator}
&\Pi({\bm r}) = \frac{T m^2}{2\pi^2 rk_{\mathrm{F}}} \frac{\mathrm{csch}\{\frac{\pi T}{v_{\mathrm{F}}} r [f_{+}(\phi_r)+f_{-}(\phi_r)]\}}{(1-\beta^2m^2)\sqrt{f_{+}(\phi_r)f_{-}(\phi_r)}}
\\\nonumber
&\times \cos\bigg\{ \frac{h_z r}{v_{\mathrm{F}}} [f_{+}(\phi_r)+ f_{-}(\phi_r)] + k_{\mathrm{F}} r [f_{-}(\phi_r)-f_{+}(\phi_r)] \bigg\},~~
\end{align}
where $v_{\mathrm{F}} = k_{\mathrm{F}} /m$ is the Fermi velocity, at $m|\beta| < 1$, expression on the second line might be estimated as $\propto \cos\left\{ \frac{2 r}{v_{\mathrm{F}}}[h_z+\beta m \mu \sin(2\phi_r)]\right\}$.
At finite exchange interactions, the paring correlations decay and oscillate in space, as expected for systems with magnetic Cooper pair-breaking source. Thus, it can emerge via $0$-$\pi$ transitions in the Josephson junctions through the d-wave AFM, as demonstrated in Ref. \cite{Linder_PRL2023, Neupert_arxiv2023, Beenakker_PRB2023}.

Next, it is instructive to comment on the superconducting transition temperature of the intrinsic superconductivity. At $\Delta \rightarrow 0$, assuming superconductivity in the spatially homogeneous regime, the BCS transition temperature can be found from the solution of the self-consistently equation
\begin{align}\label{BCS_meanfield}
\ln\frac{T}{T_{c0}}=\Psi\left(\frac{1}{2}\right) - \mathrm{Re}\left\langle\Psi_0\left(\frac{1}{2} - i \frac{h_z+m\beta \mu \sin(\phi)}{2\pi T}\right) \right\rangle,
\end{align} 
where, $\Psi_0(x)$ is the polygamma function, $\langle f(\phi)\rangle \equiv \int_0^{2\pi}\frac{d\phi}{2\pi}f(\phi)$ denotes integration over the directions of momentum and $T_{c0}$ is the transition temperature at $h_z=0$ and $\beta =0$. 

It is evident that both $h_z$ and $\beta$ terms suppress the transition temperature, in accord with the spatial oscillatory dependence of the correlator (\ref{correlator}). However, they may partially compensate each other giving rise to a residual superconducting state at larger $h_z$ at small $T$ regime, as illustrated in Fig. (\ref{fig2}). 

One might argue that the strength of d-wave AFM order can depend on the position of the chemical potential, $\beta(\mu)$. In this situation, the superconducting transition temperature can exhibit unusual behaviour as a function of electron density in thin films. 

Furthermore, it can be shown that the momentum dependence of the $\beta$-term suppresses the realization of inhomogeneous bulk Larkin - Ovchinnikov - Fulde - Ferrell state compared in contrast to the effect of $h_z$-term contribution. However, in finite-size systems with lengths on the order of several $v_{\mathrm{F}}/|\beta m \mu| \sim 1/|\beta m k_{\mathrm{F}}|$, one might expect the stabilization of such inhomogeneous state. Additionally, the inhomogeneous state might be stabilized in a d-wave superconductor when brought in contact with the d-wave AFM under certain symmetry-matching conditions, \cite{Schaffer_arxiv2023}. 
This prediction can be tested, for example, via the anomalous Little-Parks oscillations \cite{Zyuzin_JETP_2008, Zyuzin_2009}. After this general introduction, let us now investigate the magnetoelectric effect in the system.

\section{Magnetoelectric effect}
Consider a simple model to demonstrate the magnetoelectric effect in the superconductor with d-wave exchange interaction, by setting $h=0$ and focusing on the small gap regime $|\Delta| \ll T$. It allows addressing the spin polarization in the lowest order in $\Delta$. 

By construction of the d-wave exchange field in Hamiltonian (\ref{Hamiltonian12}), the superconducting current can only induce an out-of-plane component of the spin polarization density, which in momentum representation $S_z({\bm q}) = \int d{\bm r} S_z({\bm r}) e^{-i{\bm q}\cdot {\bm r}}$ can be expressed as \cite{Edelstein_1}
\begin{eqnarray}\label{Spin_polar_def}\nonumber
S_z({\bm q}) &=& -T\sum_{n}\int\frac{d{\bm p}d{\bm k}}{(2\pi)^4}  \mathrm{tr}_2\bigg\{ \sigma_z G\left(\omega_n, {\bm p}+{\bm k}+\frac{{\bm q}}{2}\right)
\\\nonumber
&\times& \Delta\left({\bm k}+\frac{{\bm q}}{2}\right) \sigma_yG\left(-\omega_n, {\bm p}\right)\sigma_y
\Delta^*\left({\bm k}-\frac{{\bm q}}{2}\right)
\\
&\times&  G\left(\omega_n, {\bm p}+{\bm k}-\frac{{\bm q}}{2}\right)\bigg \},
\end{eqnarray}
where the electron Green function in momentum representation is given by
\begin{eqnarray}
G\left(\omega_n, {\bm p}\right) = \frac{1}{2}\sum_{s=\pm 1} \frac{1+ s\sigma_z}{i\omega_n -\xi_{\bm p} - s[\beta p_x p_y + \gamma(p_x^2-p_y^2) ]}.
\end{eqnarray}

 %%%%%%%%%%%%%%%%%%%
\begin{figure}[t]
\centering
\includegraphics[width=7cm]{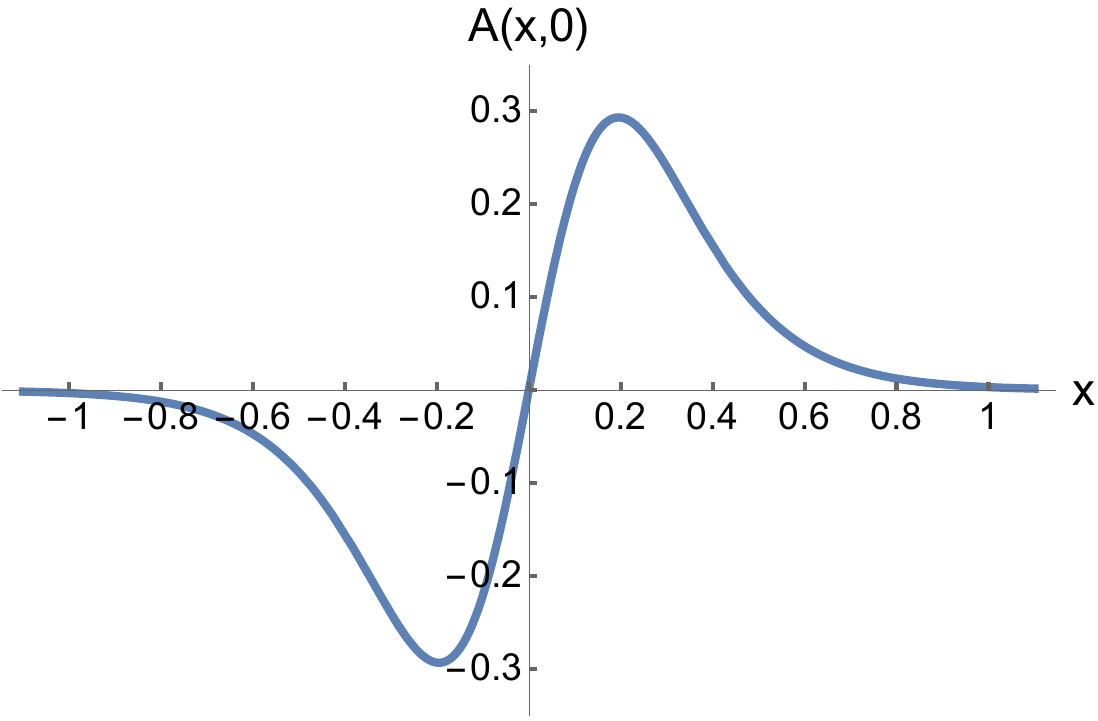}
\caption{\label{fig3} The function $\mathcal{A}(x,y)$ at $y=0$ in (\ref{Magnetoelectric_general}) of the main text. At a given $x$, the finite value of $y$ suppresses the amplitude of function $\mathcal{A}(x,y)$.}
\end{figure}
%%%%%%%%%%%%%%%%%

To proceed, we consider the long-wave limit noting $p\approx p_F \gg k, q$ and expand the Green functions $G\left(\omega_n, {\bm p}+{\bm k}\pm {\bm q}/2\right)$ in (\ref{Spin_polar_def}) over the powers of momentum ${\bm k}\pm {\bm q}/2$. 
Performing integration over ${\bm p}$ and summing over $n$, we obtain the leading contribution to the spin polarization in the lowest non-vanishing order ($m|\beta|<1, m|\gamma|<1$) in the following form
\begin{align}\nonumber
&S_z({\bm q})= -\frac{2\mu}{(2\pi T)^3} \int\frac{d{\bm k}}{(2\pi)^2} \Delta\left({\bm k}+\frac{{\bm q}}{2}\right)\Delta^*\left({\bm k}- \frac{{\bm q}}{2}\right)
\\\nonumber
&\times\bigg\{\left[3k_x k_y+ \frac{q_x q_y}{4}\right] \mathcal{A}\left( \frac{\beta m  \mu}{2\pi T}, \frac{\gamma m  \mu}{2\pi T}  \right) 
\\
&
+ \frac{1}{2}\left[3(k_x^2 -k_y^2)+ \frac{q_x^2-q_y^2}{4}\right] \mathcal{A}\left( \frac{\gamma m  \mu}{2\pi T}, \frac{\beta m  \mu}{2\pi T}  \right) 
\bigg\},
\end{align}
where
\begin{eqnarray}\label{Hamiltonian11}
\mathcal{A}(x,y)  = \frac{-\textrm{Im}}{48\pi} \left\langle \Psi_3\left[\frac{1}{2}+i(x \sin\phi+y \cos\phi)\right] \sin\phi \right\rangle
\end{eqnarray}
is a function of $x$ and $y$ as plotted in Fig. (\ref{fig3}). It is noted that the spin polarization vanishes at large $|\beta| m  \mu/T$. It can be attributed to the suppression of the supercurrent amplitude induced by the time-reversal-breaking perturbation. This stands in contrast to the Edelstein effect in superconductors with Rashba spin-orbit interaction, wherein the spin polarization increases with the increase of spin-orbit interaction strength \cite{Edelstein_1}. 

We also note that intrinsic superconductivity requires $|\beta| m  \mu/T \ll 1$, while the region of parameters $|\beta| m  \mu/T \gtrsim 1$ might be achieved in systems with proximity induced superconducting correlations.
  
Transforming the remaining integrals over momentum noting that $\int\frac{d{\bm k}}{(2\pi)^2}  [3k_xk_y+\frac{q_xq_y}{4}] \Delta\left({\bm k}+\frac{{\bm q}}{2}\right)\Delta^*\left({\bm k}-\frac{{\bm q}}{2}\right) 
= \frac{1}{2} \int d{\bm r} e^{-i{\bm q }\cdot{\bm r }}[(\partial_x\Delta)\partial_{y}\Delta^* - 2 \Delta \partial_{xy}^2\Delta^* + \mathrm{h.c.}]$,
we obtain the spin polarization density in spatial coordinate representation in the following form
\begin{align}\label{Magnetoelectric_general}
&S_z({\bm r}) =- \frac{\mu}{(2\pi T)^3} \mathcal{A}\left( \frac{\beta m  \mu}{2\pi T}, \frac{\gamma m  \mu}{2\pi T}  \right) 
\\\nonumber
&\times [(\partial_x\Delta) \partial_{y}\Delta^*  - 2\Delta \partial_{xy}^2\Delta^* +\mathrm{h.c.}] 
\\\nonumber
&-
\frac{\mu}{(2\pi T)^3} \mathcal{A}\left( \frac{\gamma m  \mu}{2\pi T}, \frac{\beta m  \mu}{2\pi T}  \right) 
\\\nonumber
&\times [|\partial_x\Delta|^2 -|\partial_y\Delta|^2 - 2\Delta (\partial_{x}^2\Delta^*-\partial_{y}^2\Delta^*) +\mathrm{h.c.}].
\end{align}

Finally, expanding in the lowest order in $\beta$ and $\gamma$ in (\ref{Magnetoelectric_general}), one obtains
$S_z({\bm r}) = - \frac{31}{32}\frac{\zeta(5)\mu^2}{\pi^4T^4} \nu \{\beta [(\partial_x\Delta) \partial_{y}\Delta^*  - 2\Delta \partial_{xy}^2\Delta^* + \mathrm{h.c.}] + \gamma[(\partial_x\Delta) \partial_{x}\Delta^* -(\partial_y\Delta) \partial_{y}\Delta^*  - 2\Delta (\partial_{x}^2\Delta^*-\partial_{y}^2\Delta^*) + \mathrm{h.c.}] \}$,
where $\nu=m/2\pi$ is the electron density of states per spin and $ 31\zeta(5)/(32 \pi^4) \approx 0.01$. It is observed that spin polarization is quadratic in the gradients of the order parameter and d-wave symmetric. 
Specifically, in the case when  $\Delta({\bm r}) = |\Delta| e^{i\phi({\bm r })}$, we find 
\begin{eqnarray}
S_z({\bm r})  \propto  -\beta(\partial_x\phi) \partial_{y}\phi 
- \gamma \left[(\partial_x\phi)^2-(\partial_y\phi)^2\right].
\end{eqnarray}
To further explore magnetoelectrics, one might recall the inverse Edelstein effect in non-centrosymmetric systems: a supercurrent diode-like response induced by the constant magnetization in combination with the spin-orbit coupling. 

In our system, however, the $h_z$-term gives rise to the transverse supercurrent component.  To demonstrate this, it is convenient to examine the Ginzburg-Landau functional density for the order parameter. 
Keeping $ {\bm h } = (0,0,h_z)$, in the lowest order in powers of $|\beta| m < 1$, $|\gamma| m < 1$ and $h_z< T$, we obtain a quadratic part in the form
\begin{eqnarray}\nonumber
\mathcal{F}_{\mathrm{GL}} &=& a|\Delta|^2 + (b-b_1) |\partial_x\Delta|^2 +(b + b_1) |\partial_y\Delta|^2 
\\
&-& b_2[(\partial_x\Delta) \partial_y \Delta^* + \mathrm{h.c.}],
\end{eqnarray}
where $a= \nu(T-T_c)/T_c$ with $T_c$ determined by Eq. (\ref{BCS_meanfield}), $b =  \frac{7\zeta(3)}{32 \pi^2} \frac{ v_{\mathrm{F}}^2}{T^2} \nu$ and $b_{1,2} =  \frac{93 \zeta(5)  }{32 \pi^4} \frac{ v_{\mathrm{F}}^2 \mu h_z}{T^4} \nu m(\gamma,\beta) $ with $|b_{1,2}|/b <1$. Here we dropped unnecessary isotropic corrections to the $b$-term, which are $ \sim O(h^2_z/T^2)$ and $\sim O((m\beta \mu/T)^2)$. 

Therefore, the supercurrent exhibits a transverse contribution in the presence of an exchange field $h_z$:
\begin{eqnarray}\nonumber
{\bm J } &=& -2i e \bigg\{ [\hat{e}_y(b + b_1) - \hat{e}_xb_2] (\Delta^* \partial_y \Delta-\Delta \partial_y \Delta^*) \\
&+&[\hat{e}_x(b - b_1) - \hat{e}_yb_2] (\Delta^*\partial_x\Delta-\Delta \partial_x \Delta^*) \bigg\}.
\end{eqnarray}
where $e<0$ is the electron charge. The proposed effect is the superconducting analog of the anisotropic linear magnetoconductivity in metallic collinear AFM investigated recently in \cite{ZyuzinVova_arxiv2023}.

\section{Discussion and Conclusions}
Let us briefly comment on the experimental observability of the proposed effect. To this end, we estimate the energy scale associated with the superconducting phase transition in the presence of the d-wave field. 
Taking $\mu\sim 1 \mathrm{eV}$ and $T_c\sim 1\mathrm{K}$, one estimates $\mu/T_c \sim 10^4$, so that for $m |\beta| \mu/T_c \sim 1$, one has to require $m |\beta| \sim 10^{-4}$. The typical value of AFM parameter in normal metal is $m|\beta| \sim 0.1$ \cite{Jungwirth_PRX2022}, however it is not known for the recently observed superconducting $\mathrm{RuO}_2$ films. 
Thus, exploring the magnetoelectric effect might require utilizing Josephson junctions through the d-wave AFM \cite{Beenakker_PRB2023}. 

It is also instructive to compare our result for the spin polarization density with the one in Ref. \cite{Edelstein_1}. 
In the limit $\alpha k_{\mathrm{F}}/\pi T\ll 1$ (where $\alpha$ is the parameter of Rashba spin-orbit interaction),  taking $\beta\sim \gamma$, we estimate the ratio of spin densities as $|S_{z}|/|S_{\mathrm{SOI}}| \propto (\frac{v_{\mathrm{F}}}{\alpha})^3 \frac{|\beta| m }{L k_\mathrm{F}}$, where $L$ is the typical length scale of the superconducting phase variation. For example taking $L\sim 1/|\beta m k_{\mathrm{F}}|$ to be the shortest length, \cite{Beenakker_PRB2023}, we find 
$|S_{z}|/|S_{\mathrm{SOI}}| \propto  (\beta  m)^2 (v_{\mathrm{F}}/\alpha)^3$.

To summarize, we investigated the magnetoelectric response in a centrosymmetric superconducting system in the presence of d-wave exchange interaction. 
We discussed an analog of the Edelstein effect, demonstrating that the spin-charge coupling leads to the spin polarization quadratic in the supercurrent and d-wave symmetric. 
Moreover, we showed that the supercurrent exhibits a transverse contribution in the presence of an isotropic magnetic exchange field.

Our results may be used as a starting point for further investigation of spin-charge coupling in superconducting structures with $(d,g,i)$-wave magnetizations. 
Here we have focused on the ballistic two-dimensional case. However, experimentally realistic superconducting systems are three-dimensional and inevitably involve disorder scattering processes \cite{Ruf_Nature2021}. The investigation of transport phenomena in such structures is of interest \cite{bobkova_PRB2023}, as recently demonstrated in \cite{Tokatly_Bergeret}.

\section{Acknowledgements}
We thank Jakub Tworzyd\l o and Vladimir Zyuzin for helpful discussions. We thank Tim Kokkeler, Ilya Tokatly, and Sebastian Bergeret for pointing out a typo in the first version of the manuscript. This work was performed as part of the Academy of Finland Centre of Excellence program (project 352925). We acknowledge support from the QuantERA II Programme that has received funding from the European Union’s Horizon 2020 research and innovation programme under Grant Agreement No 101017733. We thank the Pirinem School of Theoretical Physics for warm hospitality.

%%%%%%%%%%%%%%%%%%%%%%%%%%%%%%%%%%%%%%
\bibliography{altermagnet_references.bib}

\end{document}